\documentclass[12pt]{iopart}
\usepackage{graphicx}
\usepackage{color}
\usepackage{umlaut}
\usepackage[english]{babel}
\begin{document}
\title{Graphene on Si(111)7x7}

\author{O. Ochedowski$^1$, G. Begall$^1$, N. Scheuschner$^2$, M. El Kharrazi$^{1}$, J. Maultzsch$^2$ and M. Schleberger$^1$ \footnote{electronic address: marika.schleberger@uni-due.de}} 
\address{$^1$Fakult\"at f\"ur Physik and CeNIDE, Universit\"at Duisburg-Essen, D-47048 Duisburg, Germany}
\address{$^2$Institut für Festkörperphysik, Technische Universität Berlin, 10623 Berlin, Germany}

\begin{abstract}
We demonstrate that it is possible to mechanically exfoliate graphene under ultra high vacuum conditions on the 
atomically well defined surface of single crystalline silicon. The flakes are several 
hundred nanometers in lateral size and their optical contrast is very faint in agreement with 
calculated data. Single layer graphene is investigated by Raman mapping. The $G$ and $2D$ peaks are shifted and narrowed 
compared to undoped graphene. With spatially resolved Kelvin probe measurements we show that this is due to $p$-type 
doping with hole densities of $n_h\simeq 6 \cdot 10^{12}$~cm$^{-2}$. The {\it in vacuo} preparation technique presented here 
should open up new possibilities to influence the properties of graphene by introducing adsorbates in a controlled way.
  
\end{abstract}

\maketitle
\section{Introduction}
Graphene with its unique electronic properties is often envisaged as the material for future field effect transistors and other electronic devices \cite{Kim.2011}. Up to now graphene of the best quality with respect to important parameters as {\it e.g.}~charge carrier mobility has been obtained by mechanical exfoliation \cite{Geim.2004,Morozov.2008,Bolotin.2008}. This method makes use of adhesive tape and is applied under ambient conditions. It is therefore not surprising that usually graphene flakes as well as devices are heavily contaminated by residual glue, adsorbates such as water, carbohydrates and photoresist residues \cite{Katsnelson.2010,Cao.2011,Lin.2012}. This is a major drawback in comparison with the ultra clean epitaxial graphene flakes grown on SiC for example \cite{Berger.2006,Emtsev.2008}. Ishigami {\it et al.}~have proposed a method for {\it in situ} cleaning of photolithografically processed devices which involves annealing in H$_2$ at 400~$^{\circ}$C \cite{Ishigami.2007} but the devices are frequently operated under ambient conditions again introducing contaminants. It has been shown already that these contaminants significantly influence the properties of graphene as they may act as electron acceptors or donors \cite{Adam.2007,Hwang.2007,Moser.2008}. Charged impurities shift the Fermi level, may cause scattering by Coulomb interaction and may also be the origin of electron-hole puddles \cite{Ponomarenko.2009}. It is therefore very important to be able to investigate the specific influence of the respective adsorbates on graphene to better understand variations in transport properties of gated devices and to develop appropriate methods for their improvement. But due to the rather arbitrary nature of the contaminants this remained impossible until now. 

In this paper we show that graphene flakes can be exfoliated directly on a crystalline Si surface (without an oxide layer) under the cleanest conditions possible allowing access to unprecedented information. In addition, it has just recently been demonstrated that graphene/silcon hybrid structures are a very promising candidate for future transistors due to the adjustable Schottky barrier between the two materials \cite{Yang.2012}. The preparation procedure used in this work is based on the ad- or cohesion of two solids by attractive forces without any glue. This technique is widely used in microelectronics \cite{Christiansen.2006}, called {\it fusion bonding} or {\it wafer direct bonding}. The idea goes back to Lord Rayleigh who investigated the adhesion of polished fused quartz samples \cite{Rayleigh.1936}. The phenomenon occurs only with nearly perfectly flat and clean surfaces. The exact nature of the bonding depends crucially on the surface cleanliness. Under ambient conditions water adlayers, carbohydrates or other surface species are present and the attractive forces are mainly van der Waals or hydrogen bonds. The bonding type can be changed to covalent bonds and thus strengthend by thermal processing. In the extreme case of ultraclean surfaces in ultra high vacuum (UHV) however, covalent bonds can form directly even at room temperature \cite{Goesele.1995}. 

\section{Experimental details}
\begin{figure}
\begin{centering}
\includegraphics[width=0.7\textwidth]{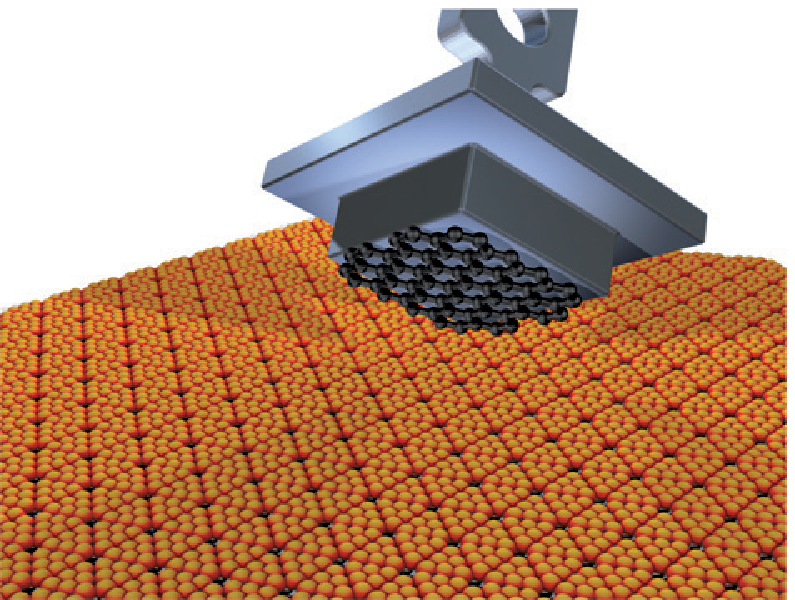}
\par\end{centering}
\caption{Schematic of the stamping procedure. The graphite flake is attached to a metal stamp which can be brought into contact with the Si(111)7x7 surface by means of a wobble stick.}
\label{Figure1}
\end{figure}

As a substrate we use a silicon wafer ($n$-doped, 10-20 $\Omega\cdot$cm) with the surface oriented perpendicular to the [111]-direction. The sample is introduced into an UHV chamber at a base pressure of $p_b\leq1 \times 10^{-10}$~mbar and degassed at $T=650~^{\circ}$C for 24 hours. The crystal is then repeatedly flash-heated up to $T=1250~^{\circ}$C for a few seconds by direct current heating. During flashing the pressure remains below $p=5 \times 10^{-9}$~mbar. This procedure removes the native oxide layer from the surface and results in the equilibrium structure of Si(111), the well known 7x7-reconstruction \cite{Takayanagi.1985}. The successfull preparation is controlled by low energy electron diffraction (LEED) which shows the typical diffraction pattern. After the substrate has cooled down again, a freshly cleaved crystal of highly oriented pyrolitic graphite (HOPG) is degassed at $T=120~^{\circ}$C and then gently pressed onto the Si surface by means of a wobblestick, as schematically shown in fig.~\ref{Figure1}. The sample is then removed from the vacuum chamber for further inspection. Note, that this was done here only to simplify the measurements in order to proof the feasibility of the deposition technique. Optical inspection and atomic force microscopy (AFM) reveal that as in the case of exfoliation under ambient conditions the stamping leads to randomly distributed flakes of graphite on the surface among which also graphene can be found, see fig.~\ref{Figure2}. 

\section{Results and Discussion}
With respect to graphene it was discussed earlier already that {\it in situ} stamping of graphene should in principle be feasible on crystalline SiO$_2$ \cite{Li.2009}. The authors calculated the energy of adhesion and cohesion, respectively, using a density functional approach. It was found that the cleavage of graphite in contact with a completely oxygen-terminated SiO$_2$ surface is very likely as it is energetically favorable. However, the experiment was performed under ambient conditions and they did not obtain single layer (SLG) but only few layer graphene (FLG). The cohesion energy of Si is typically large ($\gamma>2$~J/m$^2$) \cite{Goesele.1995} and comparable to SiO$_2$, thus one could argue along this line that the chance for SLG production should be reasonable in our case. With our approach contaminations by ambient conditions are avoided and both surfaces are very flat, thus approaching ideal conditions for fusion bonding. In addition, the Si(111)7x7 surface is known to be extremly reactive due to its specific reconstruction with unterminated bonds at the adatom positions, i.e.~with a density of one dangling bond per 5~\AA$^2$. This effect could increase the probability of covalent bond formation and may thus play an even bigger role here. 

Chen {\it et al.} used a cleaned and passivated Si surface and exfoliation in air to create a silicon/graphene device with a Schottky barrier \cite{Chen.2011}. The ideality factor of their devices was however much lower than the one achieved with the method presented in \cite{Yang.2012} indicating that the interface quality is much worse despite the clean Si surface. Ritter {\it et al.}~have applied the so-called dry contact transfer method where a braided fiberglass applicator is loaded with powder of exfoliated graphite \cite{Ritter.2008}. The applicator can be heated in UHV so that physisorbed contaminations are removed. Subsequently the applicator is brought into contact with the substrate. This procedure yields a high percentage of single layer graphene. However, the lateral dimensions of the resulting flakes is around 20 nm. Therefore, this approach produces flakes which are much too small to be investigated by means of Raman spectrosocopy and they are not suitable for device fabrication.

\begin{figure}
\begin{centering}
\includegraphics[width=0.5\paperwidth]{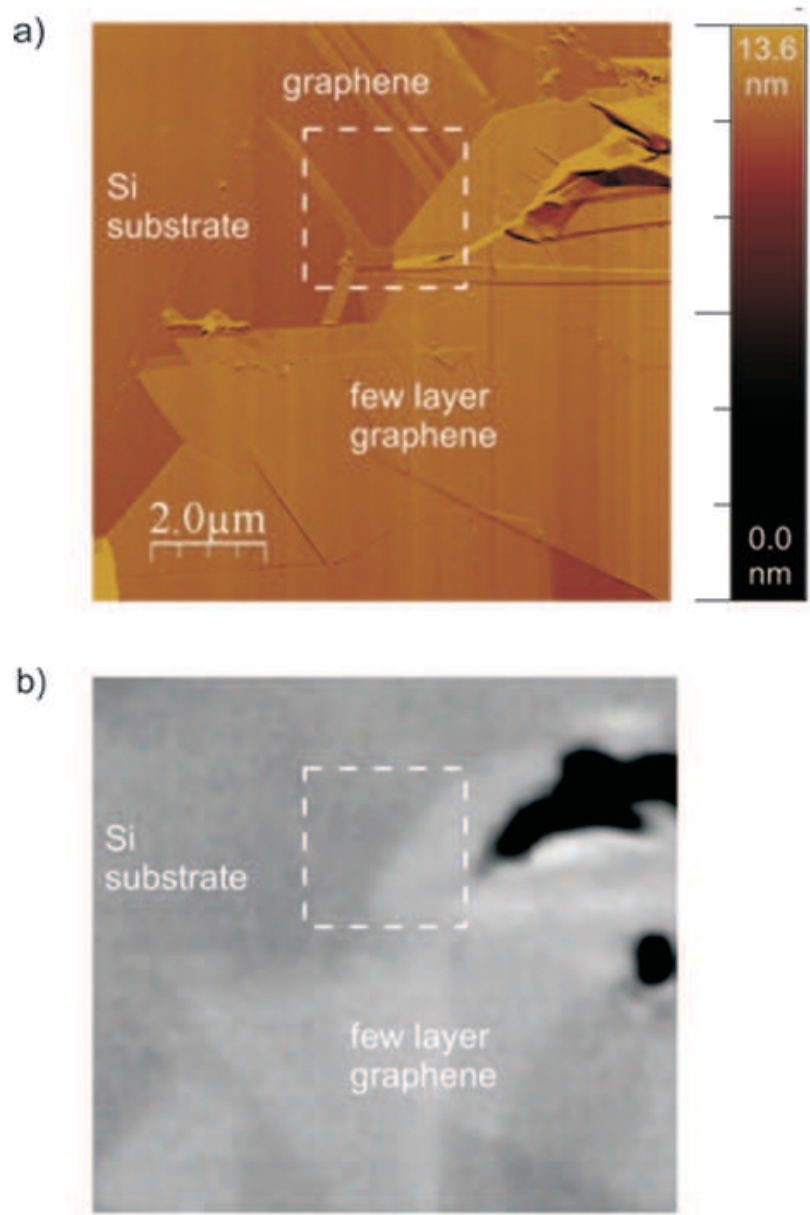}
\par\end{centering}
\caption{(a) Atomic force microscopy (tapping mode, Nanosensors NCHR with $f$=290~kHz; scan frequency 0.8 Hz) images of graphene exfoliated under UHV conditions. The graphite regions can be used to calibrate Kelvin probe data (see text). (b) Optical microscopy image of the same region as in (a). The optical contrast is very feeble and prevents easy identification of single layer graphene. Colour online.}
\label{Figure2}
\end{figure}

\begin{figure}
\begin{centering}
\includegraphics[width=0.7\paperwidth]{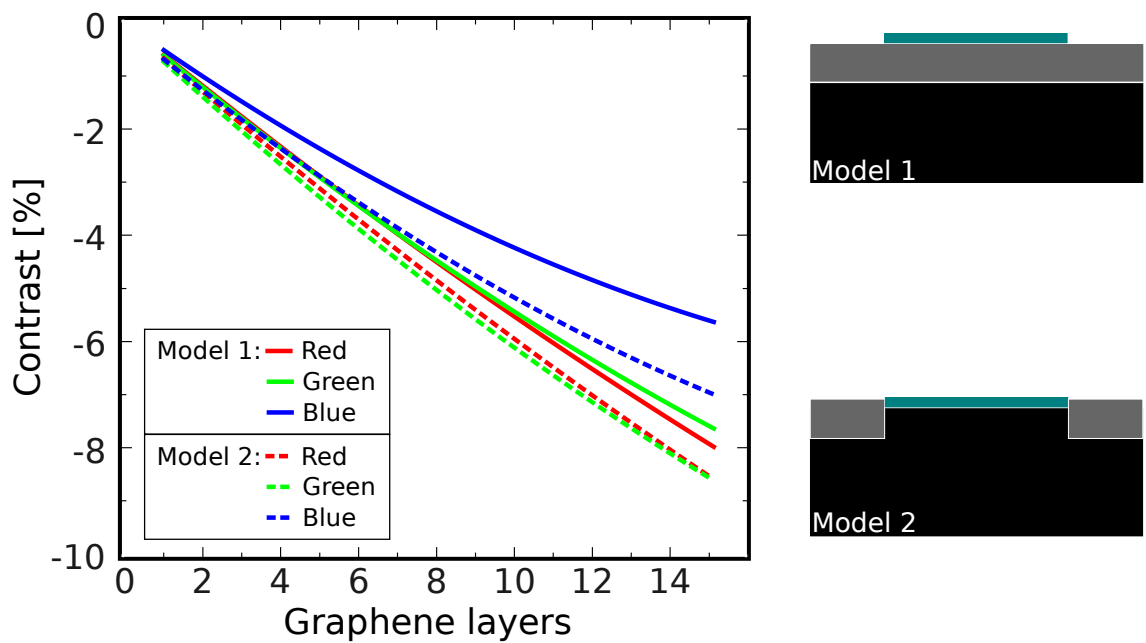}
\par\end{centering}
\caption{Calculated contrast values $C$ in \% for varying number of graphene layers and different colour channels in the case of oxidation of Si underneath graphene (model 1, solid lines) and complete protection from oxidation of Si by graphene (model 2, dashed lines). Colour online.}
\label{Figure3}
\end{figure}

This is different with the technique presented here. Typical images from stamped graphene on Si(111) taken with an optical microscope are shown in fig.~\ref{Figure2}(b). The flake distribution and size resembles the one typically found with exfoliation under ambient conditions on various substrates \cite{Akcoeltekin.2009}. Graphene flakes appear brighter than the substrate but the contrast is very faint, i.e.~$C\approx-10\pm7\%$ for 7 layers, $C\approx-6\pm6\%$ for 4 layers, and $C\leq-1\%$ for SLG. These values have the right sign (flakes appear brighter than the substrate) and are somewhat higher than the calculated data using a Fresnel law based model \cite{Blake.2007} (see below and fig.~\ref{Figure3}).

\begin{figure}
\begin{centering}
\includegraphics[width=0.55\paperwidth]{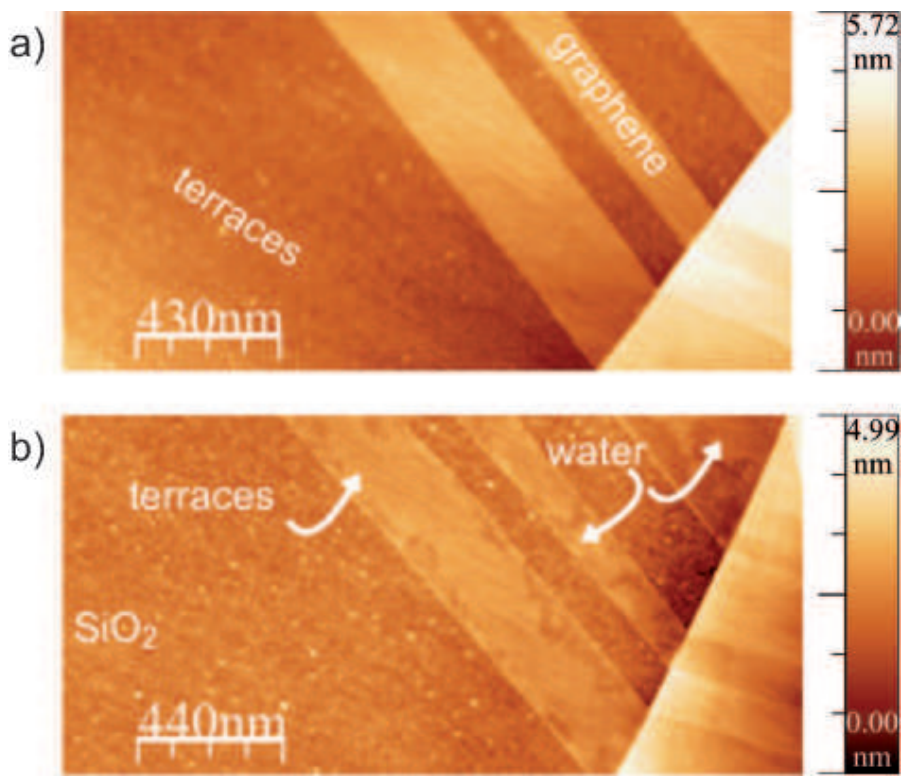}
\par\end{centering}
\caption{AFM image of a substrate region covered with graphene. The original terrace morphology of the subsrate is preserved (see text). Colour online.}
\label{AFM2}
\end{figure}

Due to the exposure to ambient conditions the substrate is covered by a native oxide layer of $d_{SiO2}\approx1.5$ nm thickness. A zoom-in (see \fref{AFM2}) reveals that the subtrate below the graphene even after extendend exposure to air still exhibits the original terrace structure of the Si substrate (faint diagonal lines running from the upper left to the lower right) which is no longer present in substrate areas not covered by graphene. It has recently been shown that graphene protects the underlying surface quite well even to the extreme of preserving the very sensitive surface state of Ir(111) under ambient conditions \cite{Varykhalov.2012}. Therefore, the 7x7-reconstruction might still be present underneath graphene. To check whether optical data can provide the answer to this question we have to calculated the optical contrast $C=\frac{R_0-R}{R_0}$, with $R$ the reflected intensity with and $R_0$ the reflected intensity without graphene. We assumed full oxidation underneath the graphene layer (model 1) and complete protection by graphene (model 2), respectively. From fig.~\ref{Figure3} one can see that the absolute contrast values are in general higher for model 2. However, the maximum difference found in the area shown in fig.~\ref{Figure2} of $\Delta C=C_{model1}-C_{model2}=1.3$~\% is clearly beyond our experimental resolution. The flakes are in principle large enough to be investigated with $\mu$-LEED which could resolve this issue. 

\begin{figure}
\begin{centering}
\includegraphics[width=0.4\paperwidth]{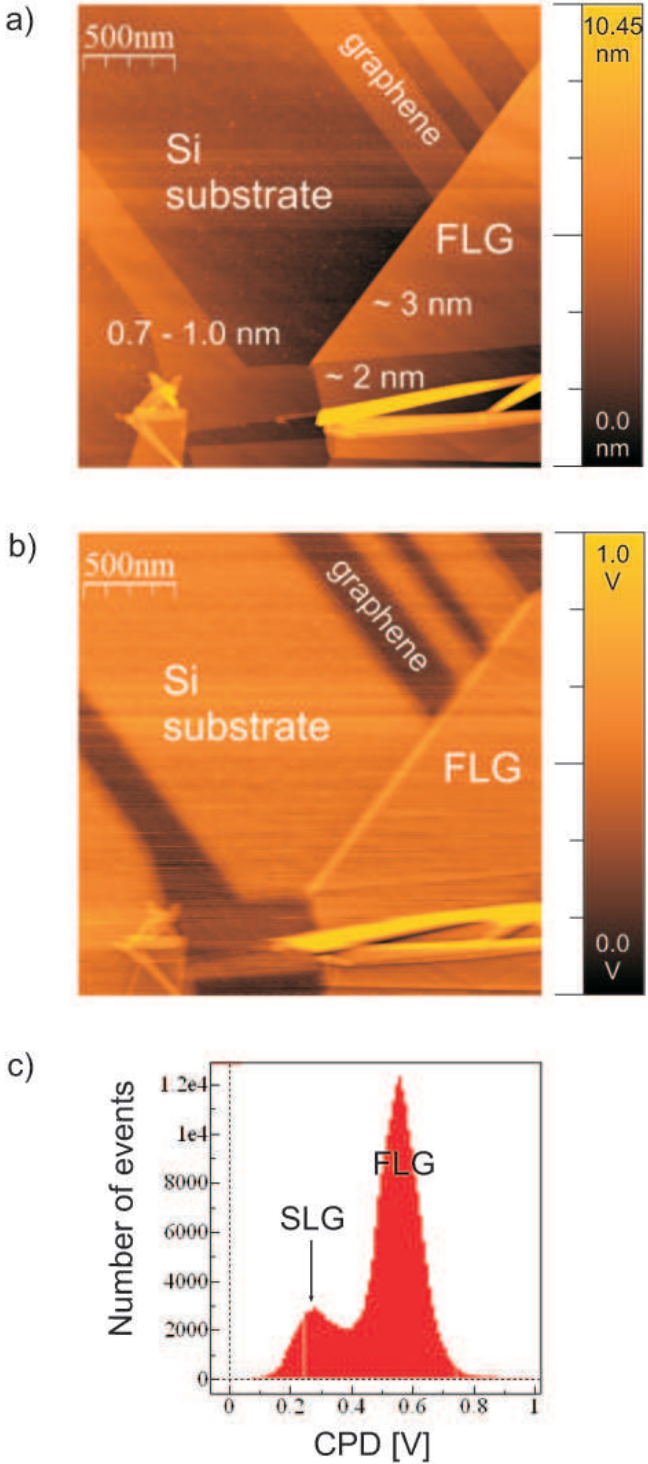}
\par\end{centering}
\caption{(a)AMF image of region marked in fig.~\ref{Figure2} where typical heights of SLG and FLG with respect to the substrate can be seen. (b) LCPD image of SLG and substrate as well as FLG (region marked in fig.~\ref{Figure2}) obtained by Kelvin probe microscopy. Bias voltage was applied to the tip. Single layer graphene was verified by Raman spectroscopy. (c) CPD histogram from (b). Colour online. }
\label{Figure4}
\end{figure}

From our AFM data (Veeco Dimension 3100, see fig.~\ref{Figure4}) we find a minimum average height of graphene of $0.7 - 1$~~nm, which would be in good agreement with either single or bilayer graphene assuming an interlayer spacing of graphite of 3.35 \AA. We also find layers with 2~nm and 3~nm height with respect to the substrate. It is very well known that height measurements of graphene with tapping mode AFM are not unambiguous \cite{Nemesincze.2008}. Here, the post-oxidization of our sample yields an additional uncertainty. 

\begin{figure}
\begin{centering}
\includegraphics[width=0.48\paperwidth]{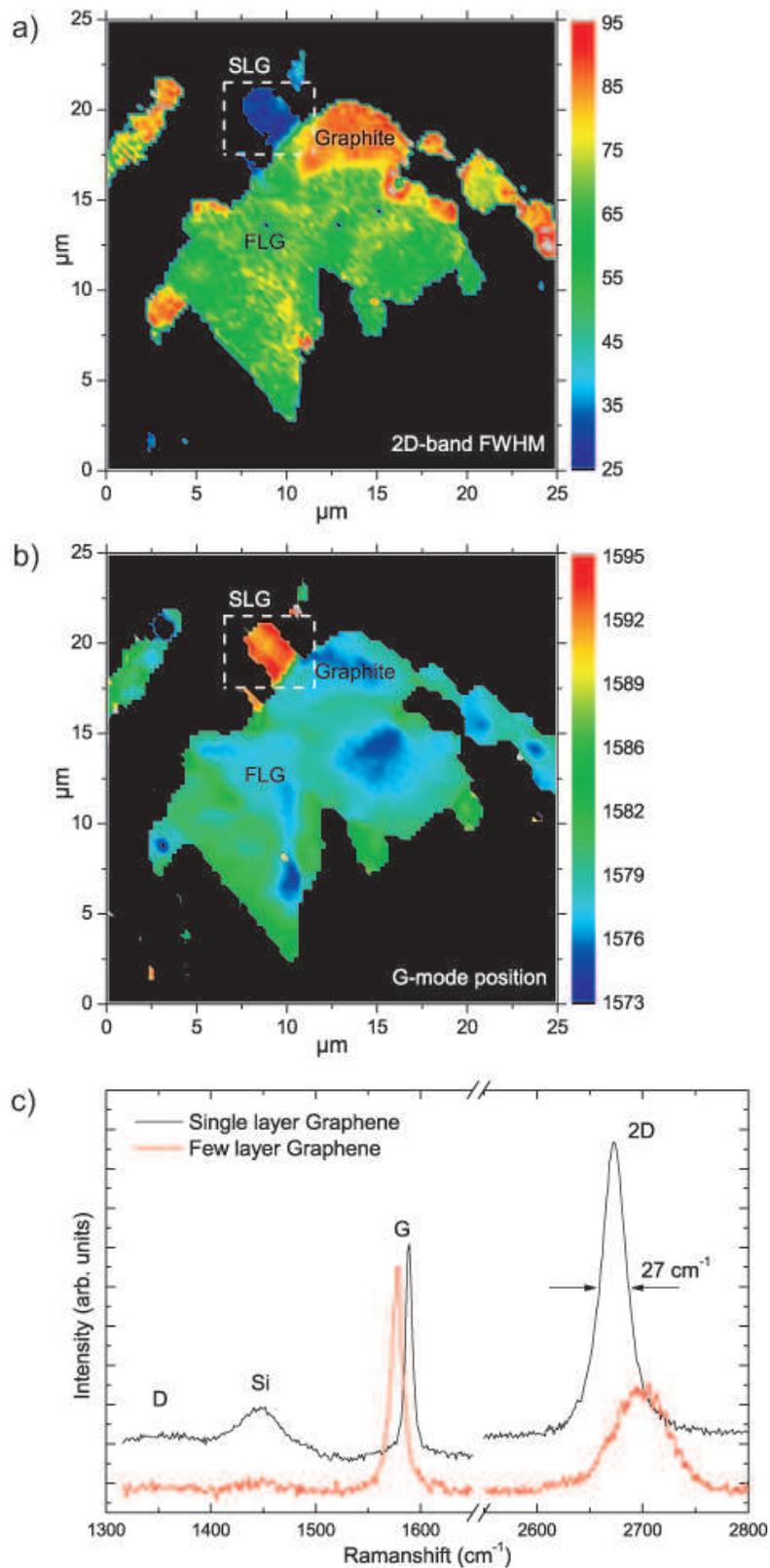}
\par\end{centering}
\caption{(a) Raman 2$D$ FWHM map of the sample region shown in fig.~\ref{Figure2}. (b) Raman $G$ mode map of the sample region shown in fig.~\ref{Figure2}. (c) Shape and width of the Raman $2D$ mode at 2675 cm$^{-1}$ are characteristic for single layer graphene. Shift and narrowing of the $G$ mode indicate doping. Colour online.}
\label{Raman}
\end{figure}

We therefore used $\mu$-Raman spectroscopy (LabRAM HR, Horiba Jobin Yvon) with an excitation wavelength of $\lambda=$532~nm to determine the number of layers. The incident power was kept below 5 mW to prevent heating. Nevertheless, we observed the formation of water adlayers after the Raman mapping, as can bee seen in the bottom part of the zoom-in, see figure 4. The spectra were calibrated with neon lines. We performed a Raman mapping with a step size of 250 nm and a laser spot of $< 0.5~\mu$m diameter. To extract the Raman shift, intensity and a full width at half maximum (FWHM) from the data the $G$ mode and the $2D$ mode were fitted separately with a single lorentzian. From the resulting FWHM map of the $2D$ mode (see fig.~\ref{Raman}(a)) we identify an SLG region (indicated by the rectangle) as well as surrounding few layer graphene and graphite. Fig.~\ref{Raman}(c) shows Raman spectra from the SLG as well as from the FLG region. The $2D$ mode for the SLG is found at 2673 cm${^-1}$, with a FWHM of 27 cm$^{-1}$ and exhibits the narrow symmetric line shape characteristic for SLG \cite{Ferrari.2006}. The absence of the disorder inducted $D$ peak at 1350 cm$^{-1}$ indicates high structural integrity of the flakes, which is typical for exfoliated graphene.

In the SLG region the $G$ mode is upshifted up to 1593 cm$^{-1}$ and strongly narrowed (FWHM ~7 cm$^{-1}$) compared to undoped graphene, which shows a Raman shift of ~1583 cm$^{-1}$ with a FWHM of 15 cm$^{-1}$ \cite{Pisana.2007}. This is a clear evidence of doping with an estimated carrier concentration of $n \geq 4 \times 10^{12}$~cm$^{-2}$ \cite{Pisana.2007,Stampfer.2007,Yan.2007}. In the few-layer graphene regions (see fig.~\ref{Raman}(b)) the $G$ mode shows lower frequencies, indicating less effective doping in thicker layers. However, for an accurate quantification of type and value of the charge carrier concentration one would have to perform experiments with a defined gate structure.

To further investigate the doping of the graphene we measured the locally resolved contact potential difference (LCPD) between the tip and the sample with a Kelvin probe setup \cite{Nonnenmacher.1991}. Kelvin Probe measurements were performed in a two pass mode. During the first pass the topography is measured in tapping mode and during the second pass the tip is lifted by 3-10 nm.  While lifted an ac bias of about $U_{Bias}=$~0.5 -1.0 V is applied to the tip at its resonance frequency. The resulting electric force on the tip is minimized with a dc voltage that corresponds to the LCPD between tip and the measured area \cite{Jacobs.1997}. From fig.~\ref{Figure4}(b) and (c) one can clearly see, that the LCPD is decreasing with decreasing layer thickness. We attribute this to a $p$-type doping of graphene \cite{Filleter.2008,Ziegler.2011,Bussmann.2011}. 

\begin{figure}
\begin{centering}
\includegraphics[width=0.5\paperwidth]{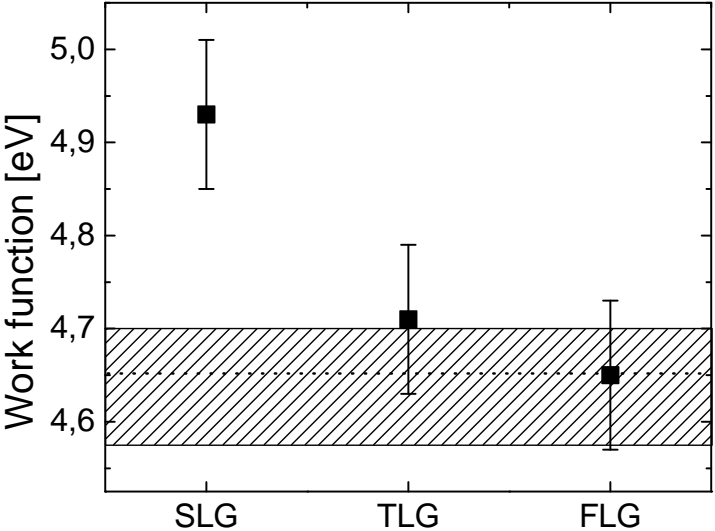}
\par\end{centering}
\caption{Work functions of single, tri- and fewlayer graphene on Si substrate determined from Kelvin probe measurements. The dashed line corresponds to the workfunction of HOPG, the hatched region corresponds to values given in the literature for free-standing (undoped) single layer graphene.}
\label{Figure6}
\end{figure}

Attributing the known work function of $\Phi_{HOPG}$=4.65~eV (see \cite{OOI.2006} and references therein) to the CPD value of the graphite regions enables us to assign work function values to our graphene layers: $\Phi_{G_i} = \Phi_{HOPG} + \Delta CPD(FLG-G_i)$, $i$ being the number of graphene layers. We can thus determine the absolute value of the work function of SLG to be $\Phi= 4.93\pm0.1$~eV (see fig.~\ref{Figure6}). For SLG, the work function variation due to doping corresponds to a shift of the Fermi energy $\Delta E_F$ with respect to the Dirac point \cite{Ziegler.2011}. The upshift of the Fermi level with respect to the value for undoped free-standing graphene $\Phi=4.57$ to 4.7~eV \cite{Sque.2007,Giovannetti.2008,Yu.2009}, is $\Delta E_{F}\approx290$~meV. This corresponds to a charge carrier density of $n_h = \frac{1}{\pi}(\frac{\Delta E_F}{\hbar v_F})^2 \simeq 6 \cdot 10^{12}$~cm$^{-2}$ if we assume $v_F = 1 \times 10^{6}$ m/s for the Fermi velocity \cite{Wang.2011}. These numbers have to be treated with great care as the exposure to ambient conditions might influence the doping level as well \cite{Sabio.2008}. Note however, that the number agrees rather well with the number obtained from the shift of the Raman $G$ mode (see fig.~\ref{Raman}). Whether the accumulation of holes observed here is indeed due to the direct interaction of graphene with the clean silicon surface needs thus to be investigated in future experiments avoiding the exposure to ambient conditions alltogether.

\section{Conclusions}
In summary, we have presented a method for the deposition of single layer graphene flakes on Si(111)7x7 under UHV conditions. As the flakes reach laterals sizes of several hundred nanometers, this technique opens up a wide range of possible experiments reaching from detailed studies of adsorbate doping and cleaning protocols to the developement of more refined stamping procedures. The latter could include e.g.~sputtered substrates, thermal processing steps or intercalated HOPG crystals to facilitate single layer exfoliation. Our approach could also help to understand the origin of the strong differences in ideality factors in current graphene/silicon devices.

\section*{Acknowledgement}
This work has been supported by the German Science Foundation (SPP 1459: Graphene and SFB 616: Energy dissipation at surfaces).

\section{References}
\bibliographystyle{unsrt}
\bibliography{Exfoliation} 
\newpage

\end{document}